\documentclass{vgtc}                          

\ifpdf
  \pdfoutput=1\relax                   
  \usepackage{graphicx}                
  \DeclareGraphicsExtensions{.pdf,.png,.jpg,.jpeg} 
\else
  \usepackage{graphicx}                
  \DeclareGraphicsExtensions{.eps}     
\fi%

\graphicspath{{figures/}{pictures/}{images/}{./}} 

\usepackage{microtype}                 
\PassOptionsToPackage{warn}{textcomp}  
\usepackage{textcomp}                  
\usepackage{mathptmx}                  
\usepackage{times}                     
\usepackage{cite}                      
\usepackage{tabu}                      
\usepackage{booktabs}                  

\usepackage[table]{xcolor}
\usepackage{float}
\usepackage{algorithm}
\usepackage[noend]{algpseudocode}
\usepackage{enumitem}
\usepackage{color}

\usepackage[table]{xcolor} 
\usepackage{booktabs}
\usepackage{tikz}

\onlineid{1015}

\vgtccategory{Research}

\vgtcinsertpkg

\newcommand{\theName}{\textit{Outliagnostics}}
\newcommand{\loo}{leave-one-out}

\title{\theName{}: Visualizing Temporal Discrepancy in Outlying Signatures of Data Entries}

\author{Vung Pham\thanks{e-mail: vung.pham@ttu.edu} %
\and Tommy Dang\thanks{e-mail: tommy.dang@ttu.edu}} %
\affiliation{\scriptsize Computer Science Department, Texas Tech University}

\teaser{
 \includegraphics[width=\linewidth]{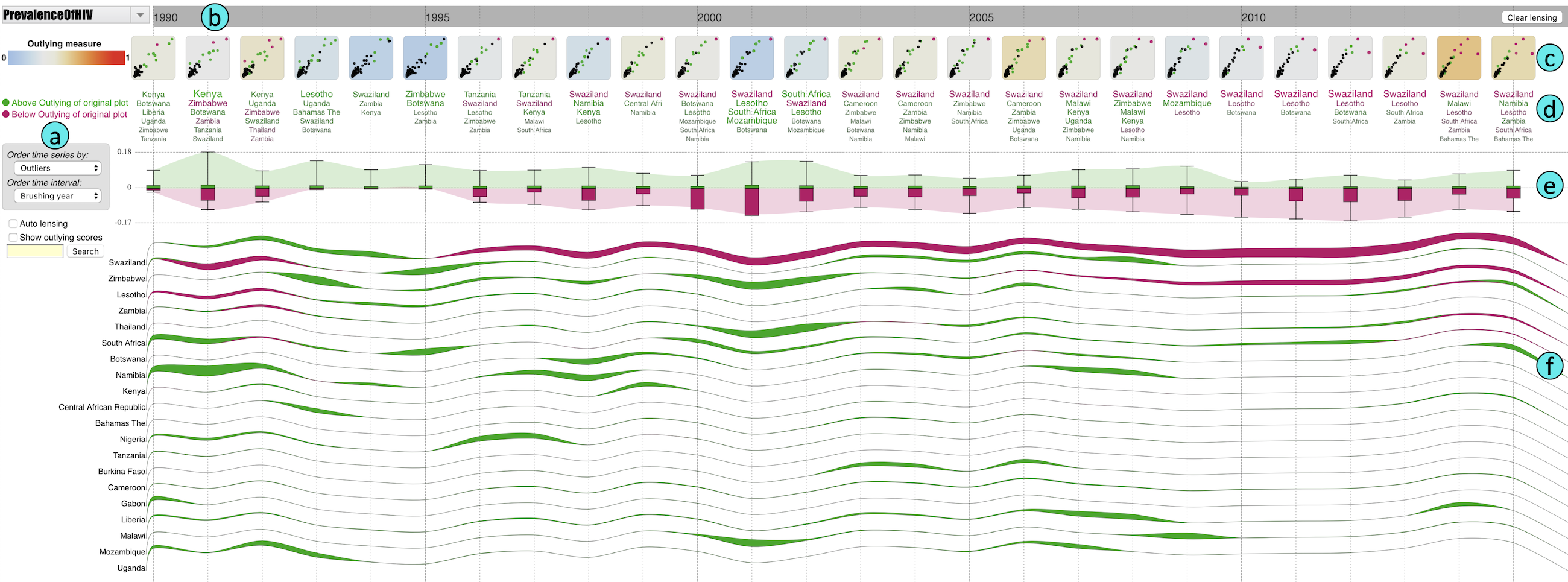}
 \centering
 \caption{Visualizing the prevalence of HIV (male vs. female) using \theName{}: (a) The control panel, (b) the lensing area, (c) the scatterplot series, (d) the top countries clouds, (e) the customized outlying boxplots, and (f) the country outlying profiles.}
 \label{fig:teaser}
}

\abstract{This paper presents an approach to analyzing two-dimensional temporal datasets focusing on identifying observations that are significant in calculating the outliers of a scatterplot. We also propose a prototype, called \theName{}, to guide users when interactively exploring abnormalities in large time series. Instead of focusing on detecting outliers at each time point, we monitor and display the discrepant temporal signatures of each data entry concerning the overall distributions. Our prototype is designed to handle these tasks in parallel to improve performance. To highlight the benefits and performance of our approach, we illustrate and validate the use of \theName{} on real-world datasets of various sizes in different parallelism configurations. This work also discusses how to extend these ideas to handle time series with a higher number of dimensions and provides a prototype for this type of datasets.%
} 

\begin{document}


\firstsection{Introduction}

\maketitle

Analyzing outliers is one of the most fundamental research areas in the field of statistics. An \textit{outlier} is an observation that appears to deviate significantly from the other observations in the sample~\cite{Hawkins1980Outliers}. Identification of potential outliers is important in many applications. Outliers may indicate errors. For example, the data may have been collected mistakenly, or an experiment may not have been set up and/or executed accurately. Outliers may be extreme cases in a distribution which are particularly interesting and important to be located.

On the other hand, an \textit{inlier} is a value that lies in the interior of a statistical distribution and practically impossible to identify~\cite{hayden2005dataset}, but in the multivariate case, thanks to interrelationships between variables, values can be identified that are observed to be more central in a distribution but would be expected to be more outlying~\cite{greenacre2014identifying}. The term ``inlier'' that we use in this paper is slightly different. We define inliers as observations that lie in the interior of statistical distribution, and their absences allow identifying outliers easier or possible. In other words, let $A$ and $B$ are two genuine observations in a distribution, but when we remove $A$, $B$ now becomes an outlier. In this case, $A$ is considered as an inlier. We use this \textit{leave-one-out} approach to measuring the significance (for both inlier and outlier) of individual data points in computing outlying as a whole. In particular, we use the Tukey outlier detection model that leverages the Interquartile range (IQR) to detect outliers in a scatterplot base on the edge lengths of the Minimum Spanning Tree (MST). The leave-one-out approach is computationally intensive, but with the use of parallel computing and binning, the time complexity of our approach is near-linear with respect to the size of a dataset. Our \loo{} approach is ``selective" since we only leave out singleton bins because removing an observation from a dense bin will not affect the outlying scores. Our contributions in this paper are:
\begin{itemize}
    \item We present an approach for measuring contributions of data points in a scatterplot outlying measure based on the ``selective'' leave-one-out cross-validation idea. The leave-one-out strategy is applied parallelly over the entire time series to formulate the outlying signatures of individual instance in the data. 
    
    \item We propose a prototype, \theName{}, to guide users on interactively exploring high dimensional datasets focusing on outliers and inliers. The visual interface supports a full range of interactions, such as lensing, brushing and linking, and filtering. The interactions and the visual interfaces are non-blocking via multithreading.
    
    \item We highlight the benefits of our approach by using \theName{} on real-world datasets. We conduct an informal study with three industry experts on real-time monitoring and detecting unusual events in High-Performance Computing center. We also present a quantitative test to evaluate the feasibility of handling large datasets on different parallelism configurations. 
\end{itemize}

The paper is structured as follows: We describe related work in the following section. Then we introduce our \theName{} prototype and illustrate it on real datasets. We present use cases and test results on running times with different parallelism settings in the Experiments section. Finally, we conclude our work in the last section.

\section{Related Work}
\label{sec:related-work}
In this section, we do not attempt to survey the full range of currently available methods. Instead, we focus on the most related techniques in discovering multivariate outliers. In particular, we review MST outliers in Section~\ref{sec:2.1} and other approaches in Section~\ref{sec:2.2}.

\subsection{The Box Plot Rule for MST outliers}
\label{sec:2.1}
We use John Tukey's method of leveraging the Interquartile range (IQR) to detect outliers in a dataset~\cite{Tukey1981Exploratory}. This method is applicable to most ranges and can be used to detect multidimensional outliers since it is not dependent on distributional assumptions. It also ignores the statistical mean and standard deviation, making it resistant to being influenced by the extreme values in the range. The interquartile range is defined as:
\begin{equation}
IQR = 3^{rd}~quartile~value - 1^{st}~quartile~value
\label{eq:e1}
\end{equation}
Tukey defines the upper and lower bounds of acceptable data as:
\begin{equation}
upperbound = 3^{rd}~quartile~value + IQR*factor
\label{eq:e2}
\end{equation}
\begin{equation}
lowerbound = 1^{st}~quartile~value - IQR*factor
\label{eq:e3}
\end{equation}
where the $factor$ is set to 1.5. There does not seem to be any statistically-driven reason Tukey uses 1.5 as a hard basis for his method (we also use 1.5 as the default setting in our method). A larger number (such as $3$) could be used to identify the ``extreme'' outliers. Values above the upper bound or below the lower bound are considered as outliers~\cite{Wilkinson2005Scagnostics}. 
Since we are looking for observations that are visually deviated from the other observations in a scatterplot, we only use \textit{upperbound} in our outlier detection algorithm. Outliers are identified by the Box Plot rule (as described above) on the MST lengths. The outlying measures the proportion of the total edge length of the minimum spanning tree accounted for by the total length of outlying edges. 
\begin{equation}
c_{outlying} = length(MST_{outliers}) / length(MST)
\end{equation}
\subsection{Other Outlier detection methods}
\label{sec:2.2}
There are many survey papers and excellent books on outlier detection written by statisticians~\cite{Hadi1993Procedures,Iglewicz1993How,Barnett1978Outliers} and computer scientists~\cite{Chandola2009ADS,Aggarwal2013Outlier}. Here we focus on multivariate outlier detection techniques. Rohlf~\cite{Rohlf1975Generalization} proposes a method of detecting outliers in multivariate data by testing the largest edge of the MST with an assumption that these edges follow a gamma distribution. Similar methods based on the MST have been proposed~\cite{Lin2008Minimum}, but they suffer the problem when variates are correlated~\cite{caroni1995rohlf}. 
Nysia et al.~\cite{george2015iterative} use an iterative \loo{} approach for outlier detection in RNA-Seq Data, but it is more about improving accuracy rather than reducing computation expense. Takuro and Akihiro~\cite{kutsuna2014outlier} propose an outlier detection method based on \loo{} density using binary decision diagrams to reduce the computation expenses. 

Another popular approach to detect multivariate outlier is based on clustering~\cite{Zahn1971Graph}. Pamula et al.~\cite{Pamula2011Clustering} apply \textit{k-means} clustering algorithm to divide the data set into clusters. The point which is lying near the centroid of the cluster is not a probable candidate for an outlier, and we can prune out such points from each cluster. Next, we calculate a distance-based outlier score for remaining points. Based on the outlier score, we declare the top $n$ points with the highest score as outliers. Jiang et al.~\cite{Jiang2001TwoPhase} propose a two-phase clustering algorithm for outliers detection using a heuristic ``if one new input pattern is far enough away from all clusters' centers, then assign it as a new cluster center''. In the first phase, the traditional \textit{k-means} algorithm groups data points in the same cluster which may be all outliers or all \textit{non-outliers}. In the second phase, an MST is constructed, and then the longest edge of this MST is removed. Data points in the small clusters (the subtree with less number of nodes) are regarded as outliers. However, most clustering-based outlier detection algorithms do not scale well to larger datasets due to the computations needed to compute cluster iteratively~\cite{Marcus2015Cluster}.

To deal with the curse of dimensionality, Wilkinson~\cite{wilkinson2017visualizing} recently proposed an algorithm, called \textit{hdoutliers}, for detecting multidimensional outliers. The algorithm is designed to be paired with visualization methods that can help an analyst explore unusual features in data. The paper also presents a thorough survey with examples on different types of outliers, such as time series outliers, ipsative outliers, text outliers, graph outliers, geographic outliers, and \textit{Scagnostics} outliers. However, none of the above approaches try to detect the temporal behaviors of each data entry with respect to the overall distributions which is significant to capture in many application domains such as terrorism or real-time monitoring health status of high-performance computing systems. Our proposed approach tries to capture these unusual behaviors (genuine observations in the previous time points can suddenly become abnormalities) using \loo{}. To overcome the time complexity, we use binning (only leave the bins with single element out) and parallel computing.

There are several works using visual interfaces and interactions to explore and validate the outlying data. To name a few, EnsembLens~\cite{xu2019ensemblelens} applies ensemble analysis; Viola~\cite{cao2018voila} is based on Canonical Polyadic (CP) decomposition methods with tensor-based anomaly analysis algorithm; TargetVue~\cite{cao2016targetvue} uses TLOF \cite{breunig2000lof}; Rclens adopts active learning algorithm to identify rare category; CVExplorer \cite{cvexplorer}, MTDES \cite{mtdes}, and TimeMatrix \cite{timematrix} use visual overview with supported interactions for discovery and exploratory of data patterns. Zhang et al.~\cite{zhang2017survey} also provided a good survey of visualization for network anomalies. Belonging to this class, our work equips users with interconnected views and interactions to explore and validate the significance of individual entries in the overall distribution. 

\section{Design Decisions}
\label{sec:designdecisions}
Our proposed approach works with scatterplots contain single outlier or multiple outliers. In case of a single outlier, by leaving the only outlier out, the outlying measure reduces significantly, so it is relatively easy to detect. Our method proves its usefulness in the case of multiple outliers (this is common in many real-world datasets), which are subject to masking and swamping effects~\cite{Acua2004AMA}.

\textbf{Masking effect:} One outlier masks the second outlier if the second outlier can be considered as an outlier only by itself, but not in the presence
of the first outlier~\cite{irad2005outlier}. In other words, after the deletion of the first outlier, the second instance emerges as an outlier.

\textbf{Swamping effect:} One outlier swamps the second observation if the latter can be considered as an outlier only under the presence of the first one~\cite{Iglewicz1982Outlier}. After the deletion of the first outlier, the second observation becomes a \textit{non-outlying} observation.

Our approach highlights both masking and swamping effects. In general, we want to detect the observations (could be outliers or inliers) which have significant contributions to the outlying computation of a pairwise projection. At the same time, we want to avoid the side effects introduced by removing an observation from the original scatterplot. The next section starts with the design motivations behind selecting to work on 2D projections and other factors that are sensitive to our choice of leave-one-out.    

\subsection{Motivations: Why 2D?}
\begin{figure}[htb]
 \centering
 \includegraphics[width=\linewidth]{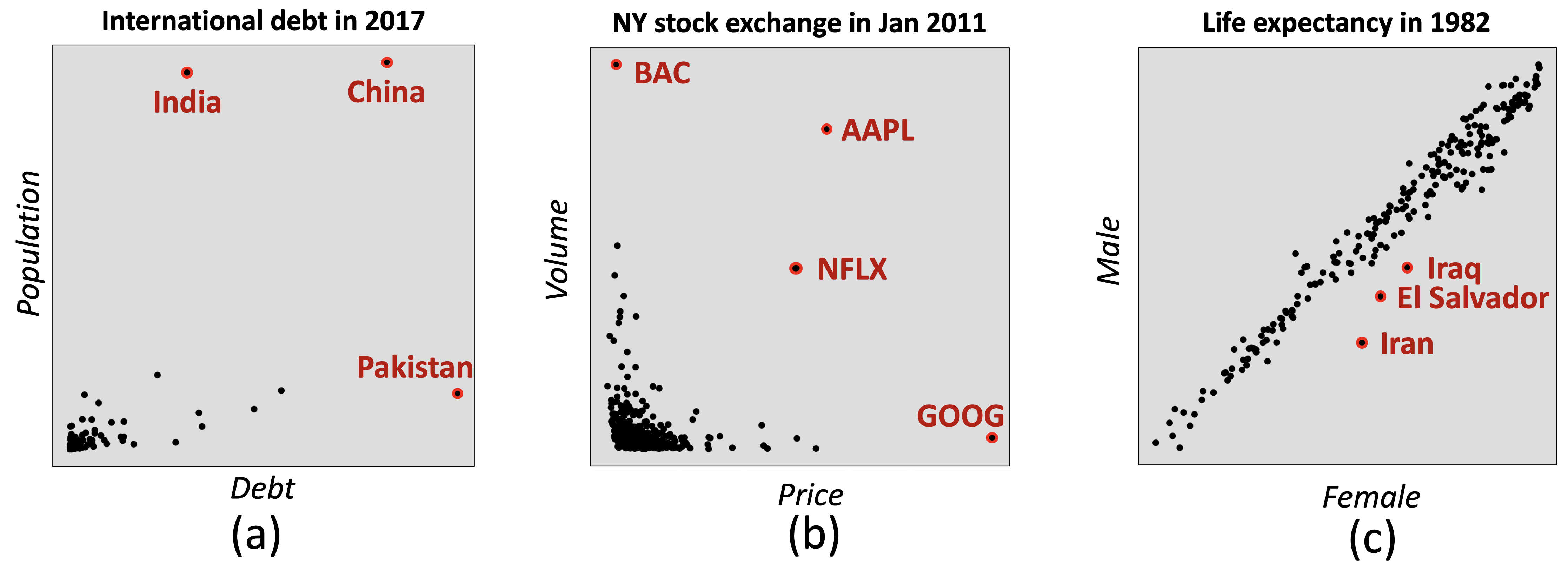}
 \caption{\label{fig:Figure3} Examples of bivariate outliers which might not be detectable in the marginal distributions:  (a) International debt, (b) New York stock exchange, and (c) Life expectancy of countries retrieved from the World Bank Database.}
\end{figure}

In many cases, multivariable data points appear to be genuine observations when each variable is considered independently~\cite{irad2005outlier}. However, a 2D projection may reveal a very different story~\cite{Jaing2001TCP}. Figure~\ref{fig:Figure3} shows an example of various cases of bivariate outliers. One might argue that bivariate outliers in Figure~\ref{fig:Figure3}(a) are detectable in the marginal distributions on $x$ axis (Pakistan), $y$ axis (India), or both (China). Nevertheless, it would be hard to refute that Netflix (NFLX) in Figure~\ref{fig:Figure3}(b) can be possibly detected as an outlier only when multivariate analysis is performed. More obviously in Figure~\ref{fig:Figure3}(c) when considering separately with respect to the spread of values along the \textit{Life expectancy of Female} and \textit{Life expectancy of Male} axes, Iraq, Iran, and El Salvador fall close to the center of the univariate distributions. The Iran-Iraq war and the Salvadoran Civil war in 1982 account for this shortage on the \textit{Life expectancy of Male} since men were needed for the wars. Thus, the test for outliers must take into account the relationships between these two variables, which in this case appear abnormal (below the diagonal of the scatterplot). Also, this work focuses on bivariate \textit{outlier/inlier} detection, but it could be generalized to work with multivariate cases.

\subsection{Design choices}
\label{sec:designchoices}
\subsubsection{Binning}
\begin{figure}[htb]
 \centering
 \includegraphics[width=\linewidth]{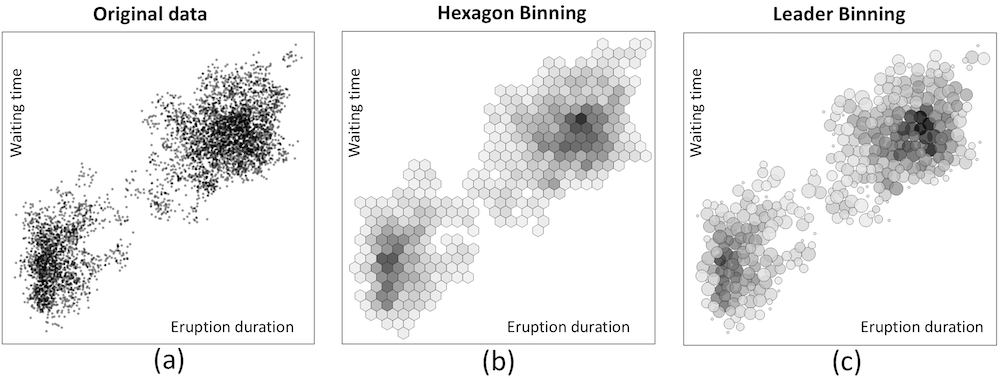}
 \caption{\label{fig:Figure4} Old Faithful Geyser data: (a) Scatterplot of \textit{eruption duration} vs. \textit{waiting time} between two consecutive eruption (b) Hexagon bins (c) Leader bins.}
\end{figure}

Before identifying outliers, we perform aggregation on the data points in scatterplots based on the following observations: 
\begin{itemize}
    \item outliers and inliers are individual observations which are distinctly distributed on a scatterplot. Binning process allows us to focus on bins with a single item. In other words, we only apply our leave-one-out approach to our singleton bins because removing an observation from a dense bin will not affect the outlying scores. This also helps to reduce the computing time significantly. We show that in Section~\ref{sec:4.4}. 
    \item We aggregate the points in each scatterplot into a certain number of bins, and then MST is computed on non-empty bins. Therefore, the complexity of our outlier detection algorithm is independent of the number of observation ($n$). Consequently, \theName{} scales well with large datasets. 
\end{itemize}
There are two standard ways to bin scatterplots: Hexagon vs. Leader algorithm~\cite{ClusteringAlgorithms}. While hexagon binning produces regular grids, leader binning starts at the positions of data points and might produce partially overlapping coverages. Both algorithms cost linear time, but we select to use leader binning since hexagon binning is sensitive to Box Plot Rule as the distance between neighboring hexagons is always the same. Moreover, while the hexagons are fixed (independent to the distribution of data points), leaders are located at the center of the clusters and hence produce smaller mean square errors~\cite{Hall1996Bin}. Figure~\ref{fig:Figure4} shows an example of hexagon binning vs. leader binning on the same input data on the left. The size of each leader indicates its coverage while the intensity of the ball highlights its density. 

Using the leader algorithm, we group data points in each scatterplot into a range from 50 to 250 clusters based on Euclidean distances among points. If there are more than 250 clusters, we increase the coverage radius and rebin. If there are less than 50 clusters, we reduce the coverage radius and rebin. The choice of coverage radius is constrained by efficiency (too many leaders slow down calculations of the Delaunay triangulation and MST) and sensitivity (too few and large clusters obscure features in the scatterplots and impact Box Plot Rule described in Section~\ref{sec:2.1}). Since we are focusing on detecting outliers (or observations that appears visually deviated from the others in a scatterplot~\cite{Hawkins1980Outliers}), increasing the number of bins to get to a reasonable representation of the scatterplot of large datasets (large $n$) does not help in highlighting outliers. In other words, we are interested in singleton and isolated bins rather than partitioning a crowded cluster into multiple smaller bins to get finer details.  

\subsubsection{Standardization for leave-one-out}
\begin{figure}[htb]
 \centering
 \includegraphics[width=\linewidth]{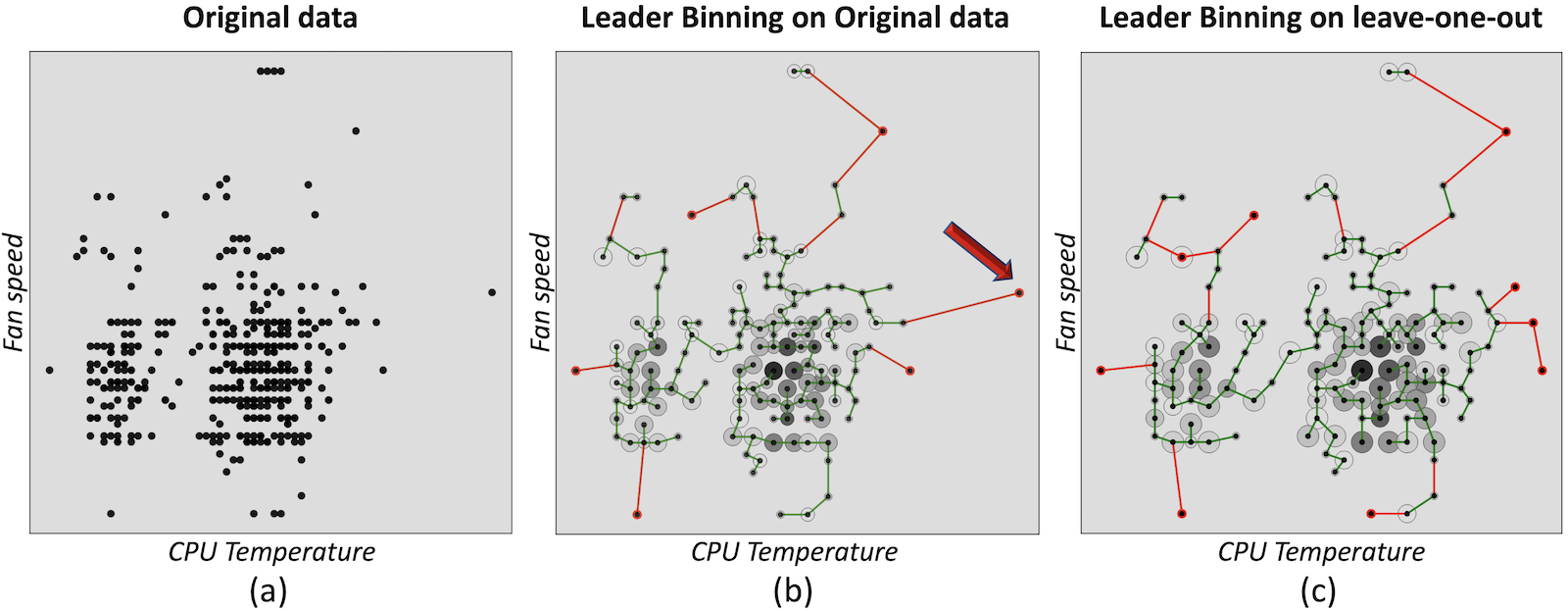}
 \caption{\label{fig:Figure5} High Performance Computing center data: (a) Scatterplot of \textit{CPU temperature} vs. \textit{fan speed}. Each data point is a computer (b) Leader bins and MST of the original scatterplots (c) Leader bins and MST of the leave-one-out plot. }
\end{figure}

When leave-one-out is applied, we do not re-standardize the remaining data points in the scatterplots. With a \textit{left-out} data point, the resulting MST could be very different as depicted in Figure~\ref{fig:Figure5}. However for many cases, the effect of removing 1 data point from the scatterplot is only on a local branch. Therefore, the MST computation time can still be improved further by localizing and recomputing only the affected branches. Moreover, we keep the same \textit{upperbound} of the Box Plot rule for computing the outlying score of all leave-one-out plots for two reasons: (1) the Box Plot rule is sensitive to the number of observations ($n$) as it uses $3^{rd}~quartile~value$ and $1^{st}~quartile~value$ to compute the IQR. And (2) an outlier in a leave-one-out plot might not be an outlier in the original scatterplot if one of the bound is modified. Reusing the \textit{upperbound} on MST lengths of the original plots for computing outlying scores on leave-one-out plots makes our approach more robust and faster. Figure~\ref{fig:Figure5} shows an example of incorrect outlying results if we do not reuse the \textit{upperbound} of the original scatterplot for computing outlying scores of a leave-one-out plot. In particular, removing the outlier at the red arrow in Figure~\ref{fig:Figure5}(b) results in the different MST in Figure~\ref{fig:Figure5}(c). Notably, the new outlying score of the leave-one-out plot is even higher than the original plot (0.26 vs. 0.21) which is incorrect.  

\subsubsection{Parallel computing}
The \loo{} approach is computationally expensive, even with the use of binning to reduce the number of calculations needed. Fortunately, many current devices and operating systems, including mobile platforms (e.g., iPhone and Android) support parallel computing with multiple cores. So, the heavy calculation could be performed in parallel with greater efficiency. Also, doing long computation tasks in the background will not block the user interfaces and interactions of our visualization.

We developed our algorithms to support parallel computation of the outlying scores. As shown in Section \ref{sec:Experiments}, using concurrent calculations improves the computation time about three times or even more with higher hardware concurrency support. Benefits of parallel computation come with its cost of creation and communication overhead, as of our experiment, the number of parallel computations should be close or equal to the number of hardware concurrency support (e.g., in JavaScript it is determined as $navigator.hardwareConcurrency$).

\subsection{\theName{} algorithm} 
Algorithm~\ref{alg:1} describes how to compute the upper bound for the box plot rule.

\begin{algorithm}[ht]
\caption{Box  Plot  Rule to identify MST outliers}\label{algorithm2}
\label{alg:1} 
\begin{algorithmic}[1]
\Procedure{computeThredshold}{$mst$}
\State \textit{// sort the MST by increasing order of edge lengths}
\State sortMSTEdgeLengths($mst$)
\State $i50$ = $mst$.length / 2
\State $i25$ = $i50$ / 2
\State $i75$ = $i50$ + $i25$
\State \Return $mst$[$i75$] + 1.5 * ($mst$[$i75$] - $mst$[$i25$])
\EndProcedure
\end{algorithmic}
\end{algorithm}

In Algorithm~\ref{alg:2}, we store the  \textit{upperbound} at line 5 and use it to compute the outlying score of the original scatterplot. For other leave-one-out scatterplots, we do not need to recompute \textit{upperbound} but reuse the same \textit{upperbound} of the original scatterplot for the new outlying computation at line 12.

\begin{algorithm}
\caption{Algorithm for computing
outlying}\label{algorithm1}
\label{alg:2} 
\begin{algorithmic}[1]
\Procedure{computOutlyingScores}{$binnedData$}
\State \textit{// compute MST of the original scatterplot}
\State $mst$ = computeMST($binnedData$)
\State \textit{// compute the upperbound of MST lengths}
\State $upper$ = computeThredshold($mst$)
\State \textit{// compute outlying score of the original scatterplot}
\State $outlying$ = MSTOutliers($upper$,$mst$)

\For{\textbf{each} data point $d$}
\State  $newBinnedData$ = leave-one-out($d$, $binnedData$)
\State $newMST$ = computeMST($newBinnedData$)
\State \textit{// Reuse the upperbound of original scatterplot}
\State $newOutlying$ = MSTOutliers($upper$,$newMST$)
\EndFor
\EndProcedure
\end{algorithmic}
\end{algorithm}

\subsection{\theName{} Components}
This section explains our approach and its applications in detail. Our general approach is similar to other typical data visualization solutions starting from statistical quantifications to visualization overviews supported with interactions for exploration and details \cite{soaviz}. Figure~\ref{fig:schema} shows a schematic overview of \theName{}: 
\begin{enumerate}

\begin{figure}[!ht]
 \centering
 \includegraphics[width=0.9\linewidth]{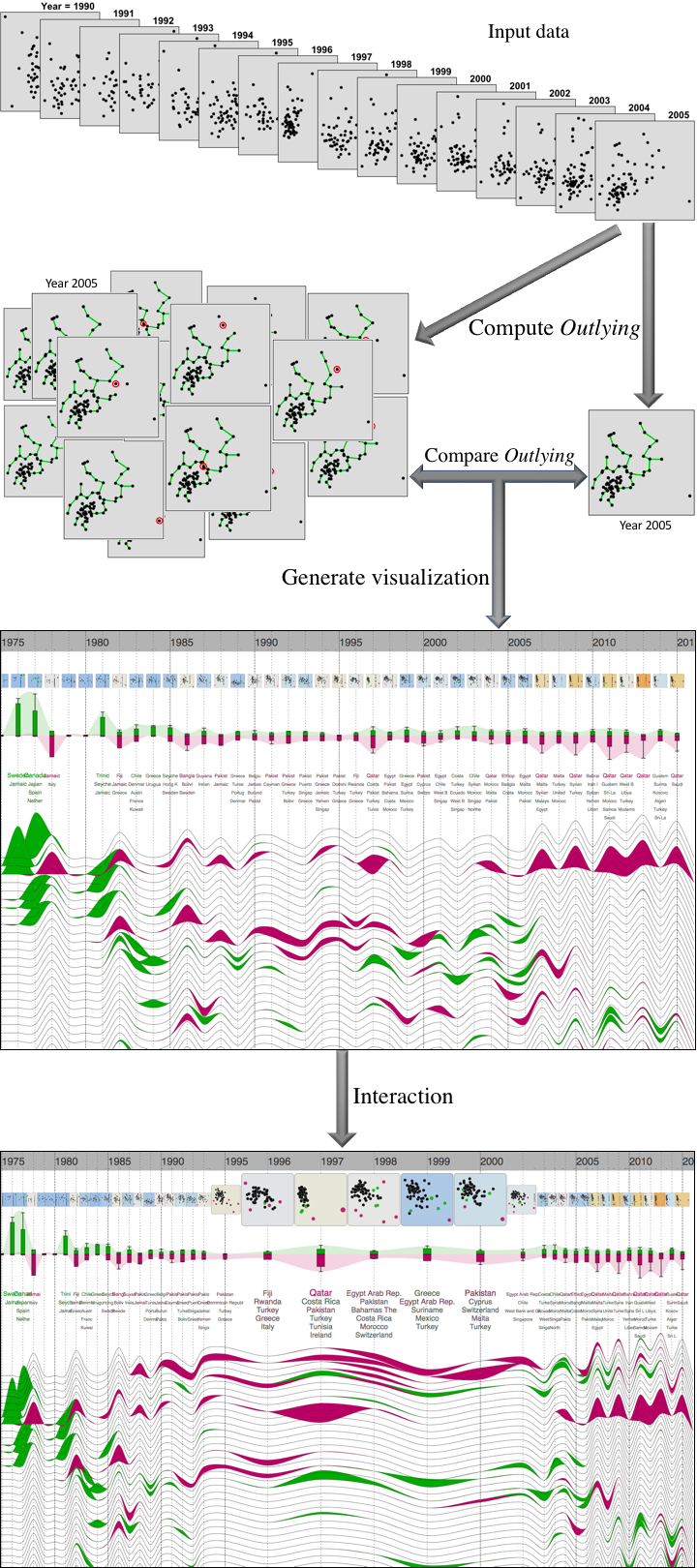}
 \caption{\label{fig:schema} A schematic overview showing the main components of \theName{}: Computing outlying scores, comparing leave-one-out outlying scores to the original scatterplots, generating the visualization, and supporting interactions. The outlying time series are color-coded based on their outlying differences when leave-one-out is applied: purple for lower while green for higher than the outlying score of original scatterplot at each time point. }
\end{figure}

\item \textbf{Processing:} Our approach computes the outlying measure of each pairwise projections in the time series. Then we repeatedly leave a data point of the plot and recompute the outlying score. Differences in outlying scores between the new and original scatterplots are recorded. We use several strategies to reduce the computation time such as binning, ``selective" \loo{}, and parallel computing as described in \ref{sec:designchoices}.

\item \textbf{Visualization:} For each variable, we display a temporal profile of outlying differences after leaving it out of the original scatterplot at each time step. We color-code observation profile. Purple is when removing the observation, the new outlying score is lower than the original scatterplot. Green is when removing the observation, the new outlying score is higher than the original scatterplot. 
\item \textbf{Interaction:} The \theName{} prototype supports filtering, ordering, brushing and linking, and lensing.
\end{enumerate}  

\begin{figure}[ht]
    \centering
    \includegraphics[width=\linewidth]{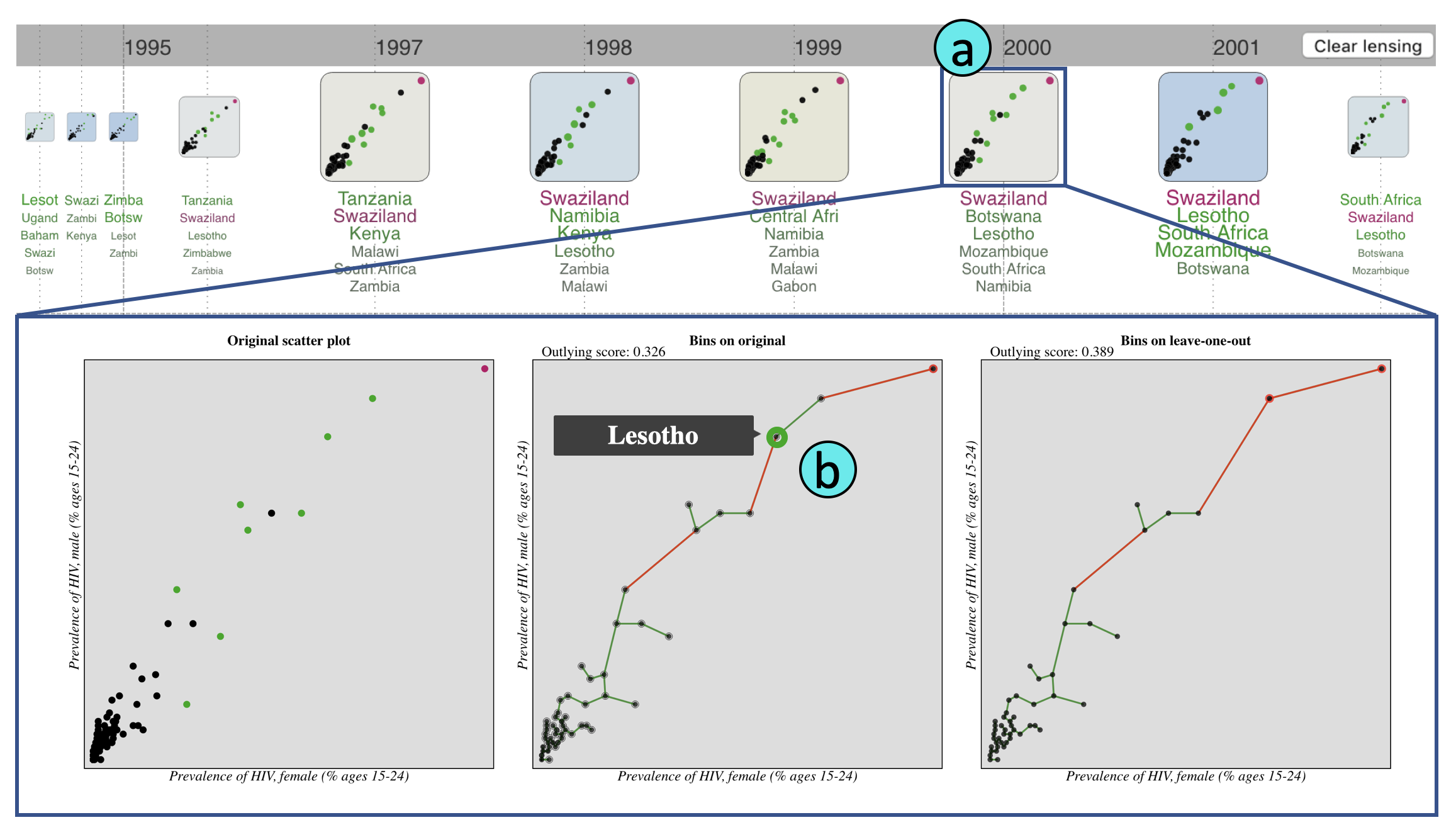}
    \caption{The lensing area, scatterplot time series, and top countries clouds for the \textit{Prevalence Of HIV} dataset.}
    \label{fig:Figure6}
\end{figure}

\textbf{Lensing area, scatterplot time series, and tag clouds:}
These are the top three components of the \theName{} interface as shown in Figure \ref{fig:teaser} and zoomed in for details in Figure \ref{fig:Figure6}. The top bar shows the timestamps where lensing is applied on mouseover, below which is the set of corresponding scatterplots: red for high outlying, blue for low outlying. This background color scheme helps users to discern the scatterplots with higher outlying scores. Below the scatterplots are the tag clouds showing the top five observations with the highest impacts (increasing or reducing) to the outlying score when leave-one-out is applied. The color of a text in these clouds also indicate if they are inliers (green) or outliers (purple).

Users can bring up a close-up window of a scatterplot via mouse click. For example in Figure~\ref{fig:Figure6}, users can click on the scatterplot for the year \textit{2000} at (a) to show details at the bottom. The first box shows the distribution of the original data; the second box shows its MST (after binning), and the third box shows the new MST for outlying calculation when leaving \textit{Lesotho} at (b) out. Red are outlying links in the MST which are longer than the \textit{upperbound} in Algorithm~\ref{alg:2}. 

\begin{figure}[ht]
    \centering
    \includegraphics[width=\linewidth]{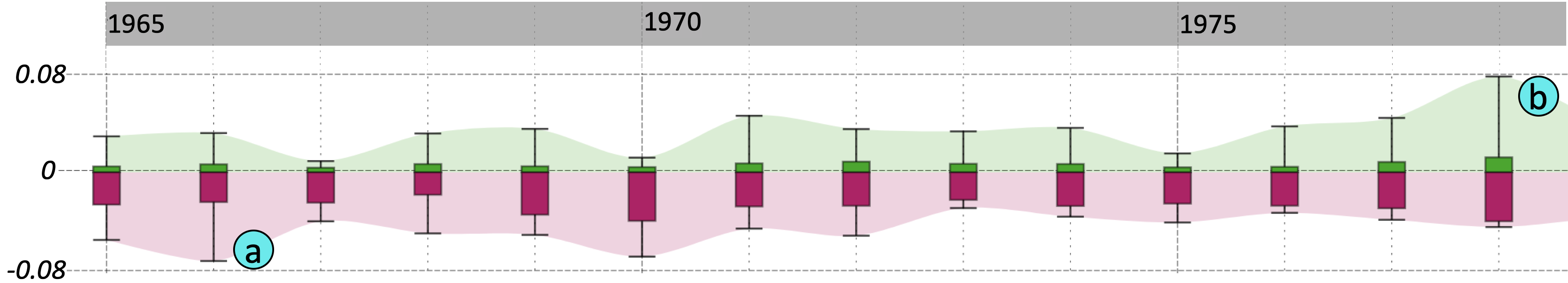}
    \caption{The customized outlying boxplots for the \textit{World Unemployment Rate} dataset.}
    \label{fig:Figure7}
\end{figure}

\textbf{The customized outlying boxplots:}
The customized outlying boxplot at each time step, as shown in Figure \ref{fig:teaser} (e) and Figure \ref{fig:Figure7}, summarizes the differences (in outlying) between the leave-one-out vs. the original plots. The zero line is the baseline for the boxplot. Above it is the green rectangle which extends from the zero line up to the average value of all the positive (inlying) differences, and the whisker at the top is the maximum of the inlying differences. On the other hand, the purple rectangle spans from the zero line down to the average of all the negative (outlying) variations, and the whisker at the bottom represents the maximum of the outlying differences. The green stream above the zero line up to the maximum inlying score and the purple one below it extending from zero line down to the maximum outlying score smooth the evolution of the inliers and outliers over time. 

Figure~\ref{fig:Figure7} shows our customized boxplots for the \textit{World Unemployment Rate} dataset. We can see that the maximum outlying difference of 0.08 (out of 1.0 as the maximum outlying score) occurred in the year 1966 (a) with several outliers such as Algeria and Oman and the maximum inlying difference of 0.08 locates in the year 1978 (b) with inliers like Lesotho, Swaziland, and Macedonia. The customized outlying boxplots and streams summarize outlying scores over time for further explorations using interactive features of the system.

\textbf{Instance profiles:}
The instance profiles allow users to dive into further details in the process of exploring outlying scores of the scatterplots overtime at the individual instance level. As depicted at the bottom panel of Figure~\ref{fig:teaser} (f), this visual area contains set of outlying time series organized in descending order based on outlying/inlying scores, depending on the user preferences by selecting the options from the control panel as in Figure~\ref{fig:teaser}(a). 

\begin{figure}[ht]
    \centering
    \includegraphics[width=\linewidth]{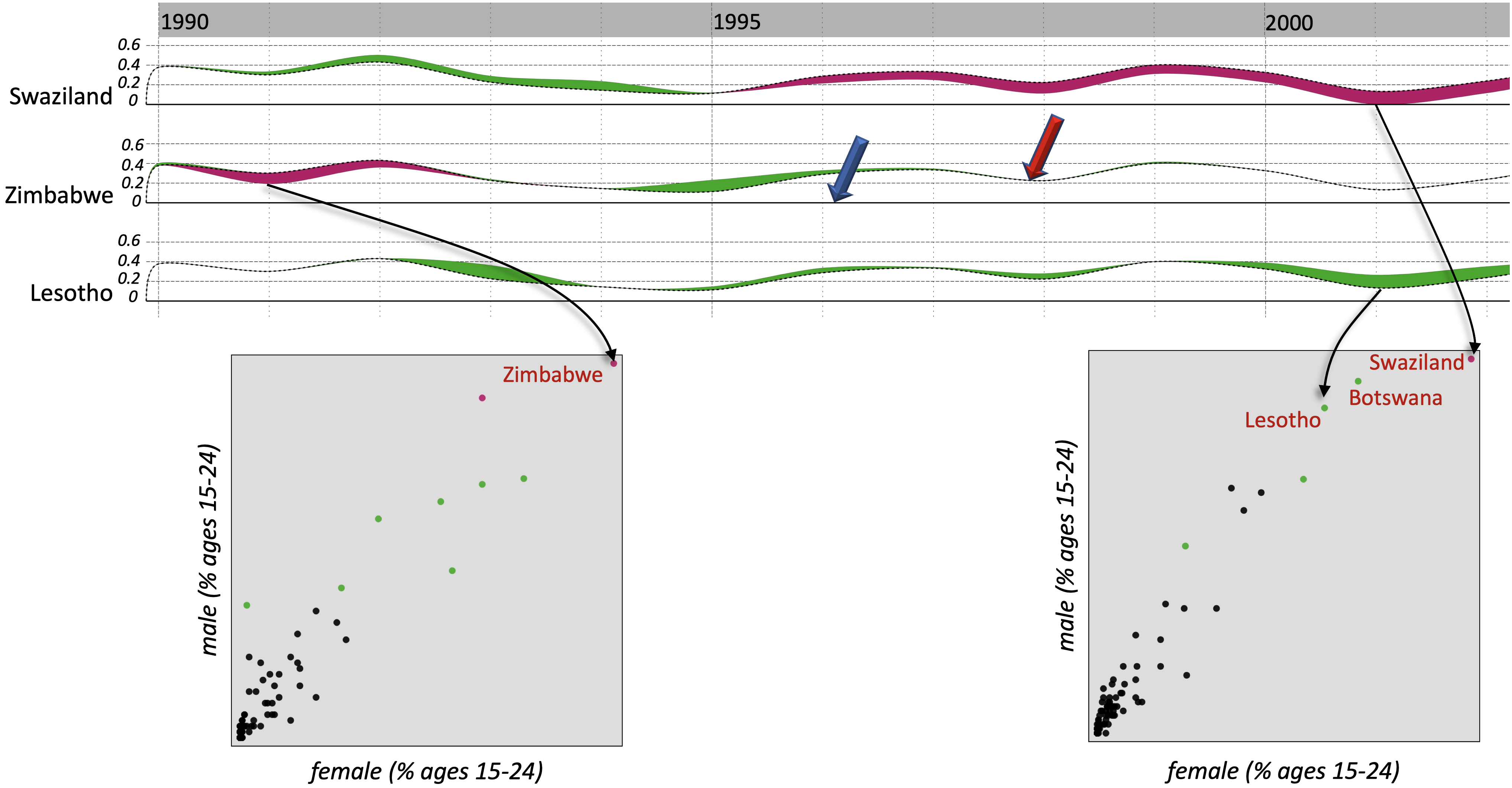}
    \caption{The profile series of three countries with high contributions to the outlying scores over years in the \textit{Prevalence of HIV} dataset.}
    \label{fig:Figure8}
\end{figure}

Figure~\ref{fig:Figure8} shows a close-up view of three countries with high impacts on the overall outlying scores overtime of the \textit{Prevalence of HIV} dataset: Swaziland, Zimbabwe, and Lesotho. Each outlying time series is projected on a baseline (the black line at the blue arrow). The dashed curve (at the red arrow) represents the outlying scores of the original scatterplots over time. The green/purple streams above/below this dashed curve represent the increments/decrements of the outlying scores when leaving the current data item out. The type of time series representation allows the users to perceive the movements of the original outlying scores quickly as well as identify the hot spots over the long time series. In particular, each row represents a unique outlying signature of the associated entry. Notice that we can filter only outlying temporal signatures with outlying/inlying scores higher than some specific threshold using a slider. Also, the profile series are automatically ordered using their overall outlying/inlying scores (when lensing is not applied) or their outlying/inlying scores at a specific time step (when lensing is applied). 

This section clarifies instance profile design via a use case, as shown in Figure \ref{fig:Figure8}. Swaziland, Zimbabwe, and Lesotho are the three countries with high impacts on the outlying score in the \textit{Prevalence of HIV} dataset. Specifically, in 1991, Zimbabwe (at the thicker purple stream) has a high outlying score impact due to its extreme values in the prevalence of HIV for both female and male (as high as 16.6\% and 6.6\%). Therefore, leaving it out will reduce the outlying score of the overall scatter plot significantly (from 0.3 down to 0.19). In 2001, Swaziland had a high impact on the outlying score due to its high values in the prevalence of HIV (30.2\% for female and 8.5\% for male). Leaving it out from this period will bring the outlying score of the profile from 0.13 down to 0. The outlying score of zero is suspicious, due to the reason that, in 2001, Botswana and Lesotho were also the two countries with high prevalences of HIV as 21.1\%, 8.0\% and 18.4\%, 7.4\% for female and male in these two countries correspondingly. Leaving Lesotho out of this profile makes Botswana become outlier (as it should be) in this time step and the outlying score increases up to 0.27. To address this, \theName{} represents Lesotho with a thicker green stream (higher inlying score) in this period, and also brings it on top of the instance profile when lensing is applied. The inlying score, in this case, is helpful in the sense that it shows the potential outlying score (0.26 instead of 0.13) of the original scatter plot. This inlying score acts as a warning to the users about the masking effect that Lesotho has on Botswana outlying status.

\begin{table*}[htb]
\caption{Prominent attributes of datasets and their example scatterplots (at the bottom) used to demonstrate our \theName{}.}
\label{table1}
     \begin{tabular}{l l l l l r r r }
   \toprule
No. &	Abbreviation &	Dataset &	Variable 1 &	Variable 2 &	Time steps &	Instances ($n$) & Singleton bins \cr
\cmidrule(lr){1-1}\cmidrule(lr){2-2}\cmidrule(l){3-3}\cmidrule(l){4-4}\cmidrule(l){5-5}\cmidrule(l){6-6}\cmidrule(l){7-7}\cmidrule(l){8-8} 
1&USUER&US Unemployment Rate&Men&Women&19&51&48\\
2&USENC&US Employment Net Change&Goods&Service&224&53&48 \\
3&WBID&International Debt Data&Total debt&Population&55&124&35 \\
4&WTRSM&World Terrorism&Attacks&Killed&48&205&39 \\
5&WBHIV&Prevalence Of HIV&Female&Male&56&217&41 \\
6&WUER&World Unemployment Rate&Female&Male&56&241&47 \\
7&WBLE&World Life Expectancy&Female &Male&56&263&27 \\
8&HPCC&High-Performance Computing&CPU Temperature&Fan Speed&33&467&63 \\
9&NYSE&New York Stock Exchange&Price&Volume&84&501&21
      \end{tabular}
\end{table*}
\renewcommand{\arraystretch}{0.05}
\begin{table*}[htb]
     \begin{tabular}{c c c c c c c c c}
      \toprule
  \raisebox{-\totalheight}{\includegraphics[width=0.084\textwidth]{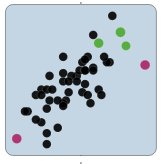}}
      &
  \raisebox{-\totalheight}{\includegraphics[width=0.084\textwidth]{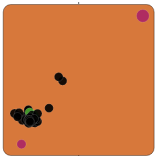}}
      &
  \raisebox{-\totalheight}{\includegraphics[width=0.084\textwidth]{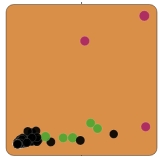}}
   &
  \raisebox{-\totalheight}{\includegraphics[width=0.084\textwidth]{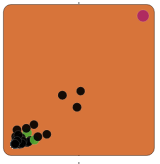}}
   &
  \raisebox{-\totalheight}{\includegraphics[width=0.084\textwidth]{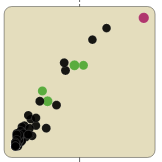}}
   &
  \raisebox{-\totalheight}{\includegraphics[width=0.084\textwidth]{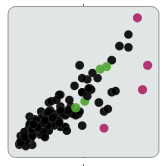}}
  &
  \raisebox{-\totalheight}{\includegraphics[width=0.084\textwidth]{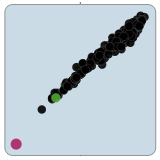}}
  &
  \raisebox{-\totalheight}{\includegraphics[width=0.084\textwidth]{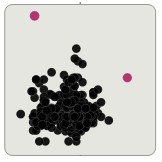}}
  &
  \raisebox{-\totalheight}{\includegraphics[width=0.084\textwidth]{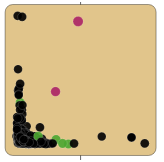}}
    \\  
USUER & USENC & WBID & WTRSM & WBHIV & WUER & WBLE & HPCC & NYSE \\ 
  \\ \bottomrule \\
     \end{tabular}
\end{table*}
\renewcommand{\arraystretch}{0.1}
\section{Experiments}
\label{sec:Experiments}
\subsection{Datasets}
\label{sec:Datasets}
We will illustrate the features and performance of \theName{} mainly through application to various datasets. Table~\ref{table1} summarizes prominent aspects of these datasets (ordered by the number of observations). The table also contains a column called \textit{Singleton bins}. These are the average actual number of times that we have to recompute MST and use Boxplot rule for detecting outliers for each dataset. As depicted, this number is independent of $n$, but depend on the shape of data distribution shown in the last row of Table~\ref{table1}.

The first two datasets are from the Bureau of Labor Statistics (BLS)~\cite{blsWebsite}. The US Unemployment Rate dataset contains the men and women unemployment rate of 51 States in 19 years from 1999 to 2017. The US Employment Net Change dataset contains all employees (in thousands, Month Net Change, seasonally adjusted) in Good Producing and Service Providing industries of 53 States from January 2000 to August 2018.

The next two datasets and the last one were retrieved from Kaggle\cite{kaggleWebsite}. The International Debt dataset from World Bank Open Data Repository\cite{wbWebsite} contains information for 124 countries from 1970 until 2024 (with projections), in particular, the debt and the population features of this dataset are used in our application. World Terrorism dataset is from National Consortium for the Study of Terrorism and Responses to Terrorism (START)\cite{startWebsite} contains terrorism data for 205 countries over 48 years (from 1970 to 2017), and we make use of the number attacks versus the number of killed variables in our use-case. The New York Stock Exchange dataset contains the information of 501 listed Stocks from January 2010 until December 2016, the price and volume dimensions of this dataset are used in our application.

The next datasets are the World Bank Open Data Repository retrieved from UIC repository\cite{UCI}. The Prevalence Of HIV dataset contains information about the prevalence of HIV female (ages 15-24) and male (in the same age range). The World Unemployment Rate dataset contains records about female and male percentage of the labor force for 241 countries over 56 years. The World Life Expectancy dataset contains male and female life expectancy of 263 countries all over the world in 56 years. To add to the variety of our datasets, we also collect health status from a High-Performance Computing Center at a university. This dataset contains fan speed and CPU temperature measurements of 467 CPUs at 33 time-steps.

\subsection{Use case 1: The Life Expectancy Dataset}
\label{sec:4.3} 
\begin{figure}[ht]
 \centering
 \includegraphics[width=\linewidth]{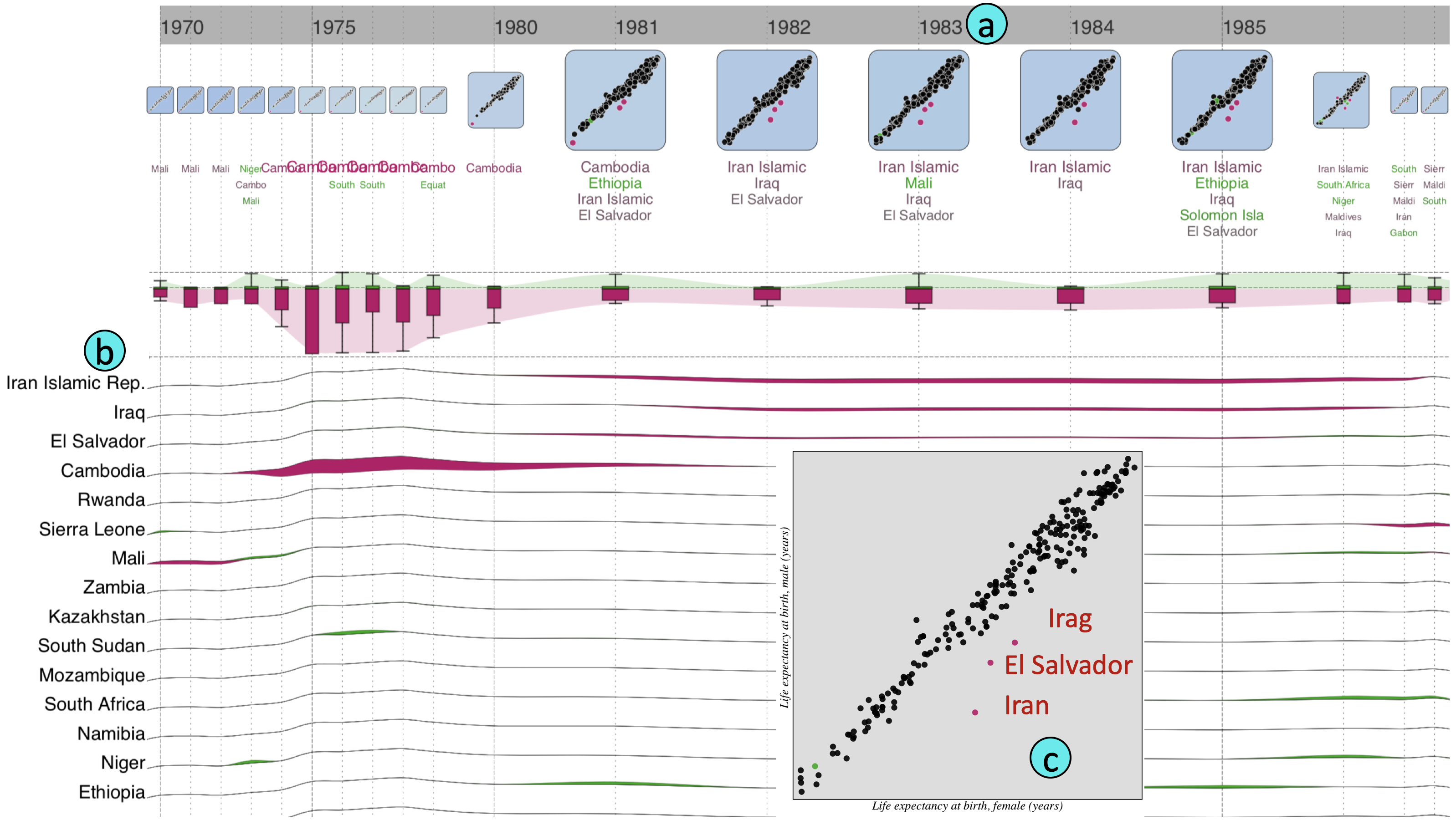}
 \caption{\label{fig:LifeExpectancyb} Lensing over the 1980s, \theName{} brings Iran, Iraq, and El Salvador on top of the item profiles section as highest contributors to the overall outlying scores in this period.}
\end{figure}

As depicted in Figure \ref{fig:LifeExpectancyb}, when lensing over the 1980s period (a), the instance profile section automatically brings Iraq, El Salvador, and Iran on top of the list (b) as these three countries had higher contributions to the overall outlying scores in this period. Clicking on a time series scatterplot in 1983 to bring the details scatterplot box (c) then mouse over Iraq, Iran, and El Salvador, \theName{} shows their corresponding life expectancy for males vs. females as 66 vs. 54, 62 vs. 47, and 63 vs. 52 correspondingly. For the Iran and Iraq case, this was due to their armed conflict from 1980 to 1988; for  El Salvador, it was the aftermath of its 1980-1992 Civil War.

This use case shows the usefulness of each component of \theName{} visualization in identifying outliers in a time series data. At the overview level, users can quickly spot out where the ``hot spots" are by viewing the fluctuation in time series boxplots to find out the places with thicker outlying/inlying streams (higher overall/potential outlying scores) to explore further. Users could further use interactive features to examine the data. For instance, users could mouse over the period with higher outlying scores in the lensing area, then clicking on the plots in the scatterplots time series section to show the \theName{} calculation details and mouse-over a data item to show its actual information at the individual data point level. Additionally, the item profile section has ordering strategies (selectable from the control panel) to bring the data items with higher contributions to the overall outlying scores to the top of the list to support further exploration.

\subsection{Use case 2: US Goods and Service Employment}
\label{sec:4.2.2} 
\begin{figure}[ht]
 \centering
 \includegraphics[width=\linewidth]{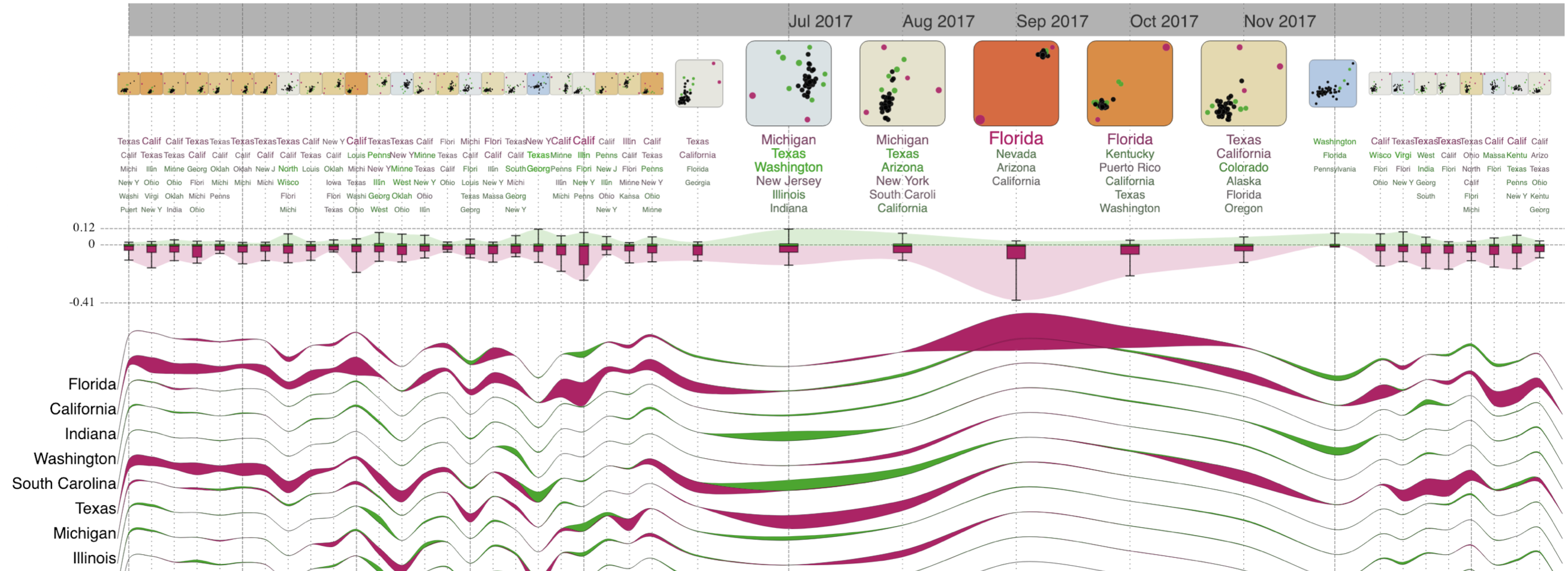}
 \caption{\label{fig:FloridaIrma} Florida as a dominant outlier in September 2017.}
\end{figure}

This use case contains US Employment Net Change data. As depicted in Figure \ref{fig:FloridaIrma}, the visual interface highlights Florida as a dominant outlier in September and October 2017 and automatically brings its outlying signature to the top when users mouse-over this period in the lensing area. These sudden net changes in employment numbers are explainable due to the impacts of the hurricane Irma in September 2017, so employment number in this city was very low both in goods-producing and service-providing in September (the change were -36,900 and -130,000 employees in goods-producing correspondingly) and got back sharply in October 2017 (+37,500 and +142,000) when the hurricane had gone.

\subsection{Use case 3: High-Performance Computing Center}
\label{sec:4.2.3} 
In this use case, we use \theName{} to monitor the health status of a High-Performance Computing Center (HPCC) at a university. In particular, the two variables being monitored are \textit{CPU temperature} and \textit{fan speed}.

\begin{figure}[ht]
 \centering
 \includegraphics[width=\linewidth]{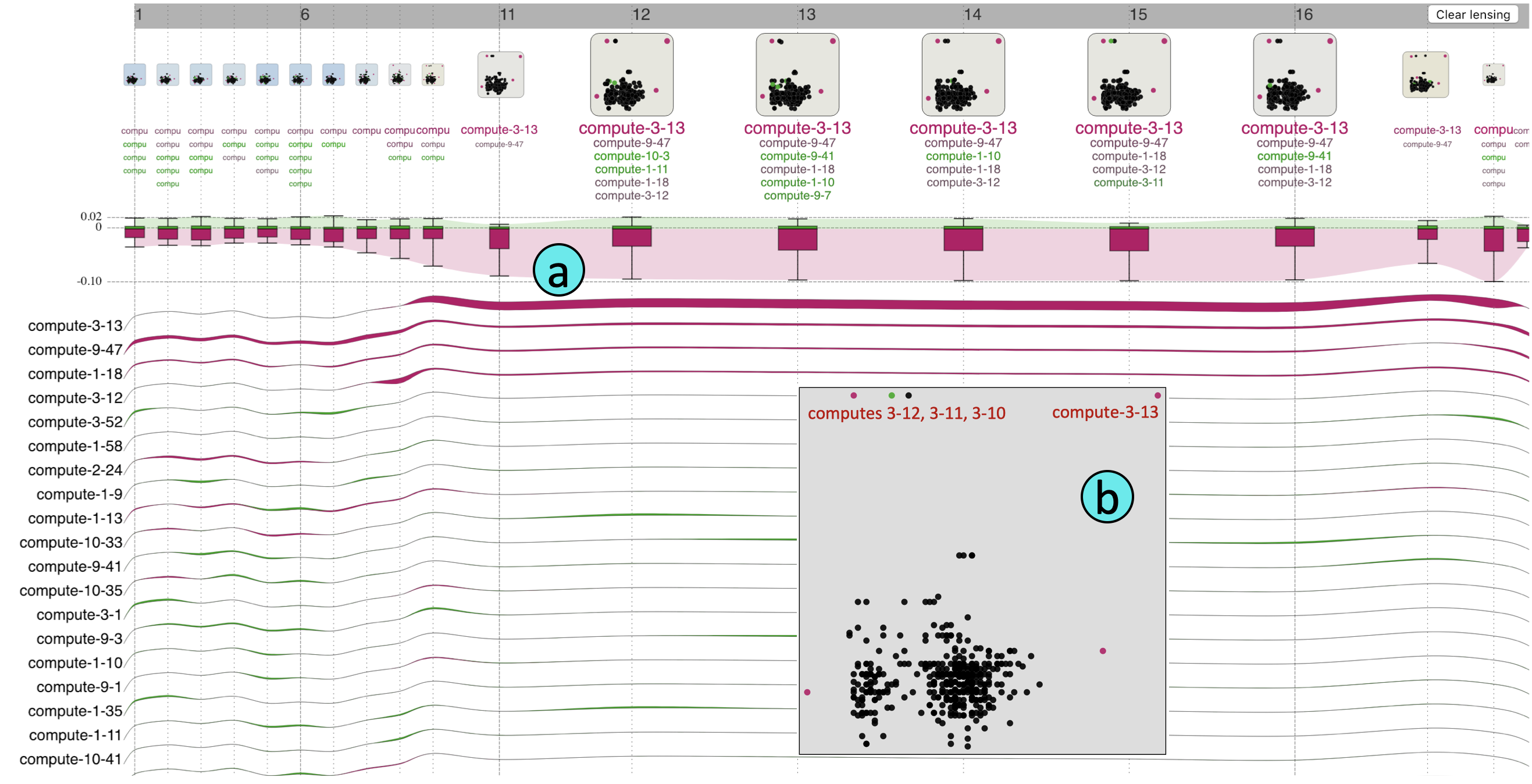}
 \caption{\label{fig:FigureHPCC2} Monitoring health status of HPCC on Wednesday, September 26, 2018: \theName{} highlights \textit{compute-3-13} experiencing overheat problem and its affect on the nearby CPUs.}
\end{figure}

Figure~\ref{fig:FigureHPCC2} shows an event at 12 PM on Wednesday, September 26, 2018: the CPU on \textit{compute-3-13} suddenly became overheated. \theName{} was able to pick up the event (a) and alerted system administrator to make CPU replacement for the malfunction CPU before it harms other neighboring CPUs. As shown in the scatterplot (b), fan speeds on \textit{compute-3-12}, \textit{compute-3-11}, and \textit{compute-3-10} had also pumped their fan rates as they sensed the heat from \textit{compute-3-13}. We discussed this thermal excursion through an informal interview with Dell experts and the HPCC director. They value the diagnostics from our prototype and suggested thermal experts and hardware team to investigate this interesting correlation between CPU temperature and fan speed. The experts commented that ``visual analytics provide an excellent opportunity to explore the correlation of hardware features'' or ``understanding the relationship of different health services is essential in our hardware design process''. \theName{} is currently deploying additional dimensions, such as real-time memory usage, power consumption, CPU load, I/O bandwidth, among other integration in this on-going collaboration.

\subsection{Implementation}
\theName{} is implemented in JavaScript using the D3.js library~\cite{bostock2011d3}. Our outlying computation and leave-one-out approach are also provided in form of JavaScript libraries. The online \theName{} prototype, demo video, source code, and more examples are available on our Github repository at \url{https://outliagnostics.github.io/}.

\subsection{Evaluation on Running Times}
\label{sec:4.4}
\subsubsection{Per scatterplot computation break-down}
In this section, we focus on evaluating the running times of \theName{} prototype on datasets of various sizes where $n$ is the number of instances (data points in each scatterplot). All tests were performed on a computer with 2.9 GHz Intel Core i5, macOS Sierra Version 10.12.1, 8 GB RAM.
Figure~\ref{fig:Eval} shows computation times broken down into the time to bin the $n$ data points, to compute Delaunay triangulation, to compute MST, and to calculate the outlying score using the Box Plot rule. In this figure, datasets are listed from left to right in the increasing order of $n$ (same order as in the Table~\ref{table1}). Here are some observations from empirical analysis:

\begin{figure}[!htb]
 \centering
 \includegraphics[width=0.93\linewidth]{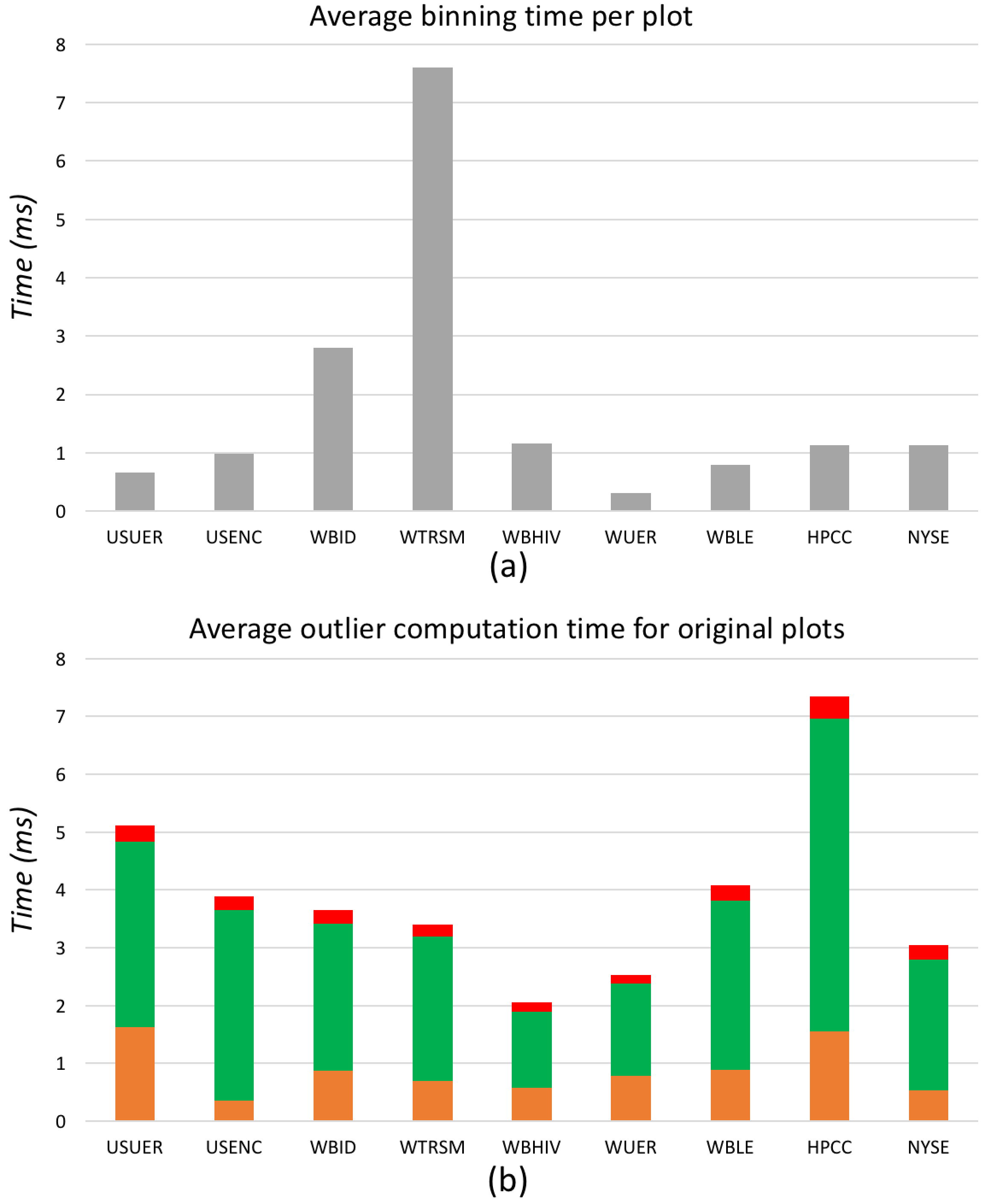}
 \includegraphics[width=0.93\linewidth]{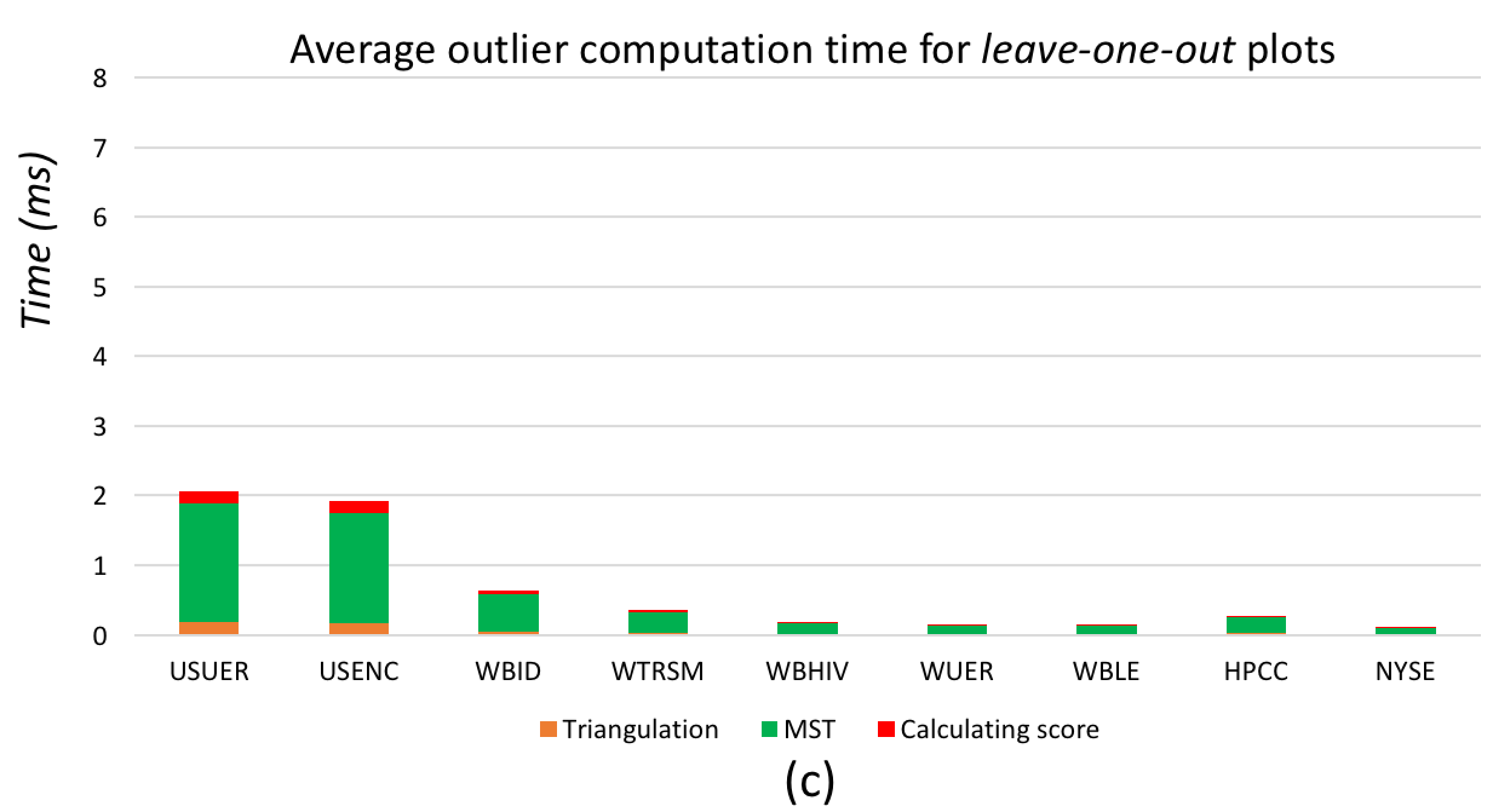}
 \caption{\label{fig:Eval} Computation times (in milliseconds) for datasets with various sizes where $n$ is the number of instances. Datasets are listed from left to right in the increasing order of $n$.}
\end{figure}

\begin{itemize}[noitemsep,nolistsep]
\item In Figure~\ref{fig:Eval}(a), WTRSM takes the most time to bin since the data distribution in a scatterplot is sparse as shown in Table~\ref{table1}. In other words, it takes a lot of time to find the right coverage radius for at least 50 leaders. In contrast, data points in USUER and WUER are well spread and take less time to come up with the number of clusters within the range from 50 to 250. Binning is done once per scatterplot (the same binning result are reused for all leave-one-out plots).
Figure~\ref{fig:Eval}(b) and (c) focus on time to compute triangulation (orange), to compute MST (green), and to calculate the outlying score (red).   
\item Figure~\ref{fig:Eval}(b) shows outlying computation time of original scatterplots, which is averaged over the entire time series. As depicted, computing MST (green) is the most expensive step while calculating the outlying score using the Box Plot rule is fast. HPPC requires the most time since most of its scatterplots are dense and hence require more time to form the MST. 

\item Figure~\ref{fig:Eval}(c) shows the average outlying computation time for leave-one-out plots. As the number of data points in scatterplots increases (going from left to right), the total computation time decreases. This is because our outlying computation time is independent of $n$ (but dependent on how the data distribution looks like: sparse or dense). Most importantly, as discussed in section \ref{sec:designdecisions}, our algorithm checks if the \textit{left-out} data point is in a singleton cluster or not. If it is, we perform the three steps in computing the outlying score: computing triangulation, forming MST, and applying the Box Plot rule. Otherwise, we skip since removing a data point from a larger cluster will not affect the final outlying score.
\end{itemize}

Overall, our proposed approach scales well with larger datasets containing thousands of observations. In other words, our leave-one-out strategy does not depend on $n$, but depends on the number of single data points (no other data point in their proximal surrounding). In fact, these are the data points that might be able to create masking and swamping effects. 

\subsubsection{Running times for different parallelizations}

\begin{figure}[!htb]
 \centering
 \includegraphics[width=0.95\linewidth]{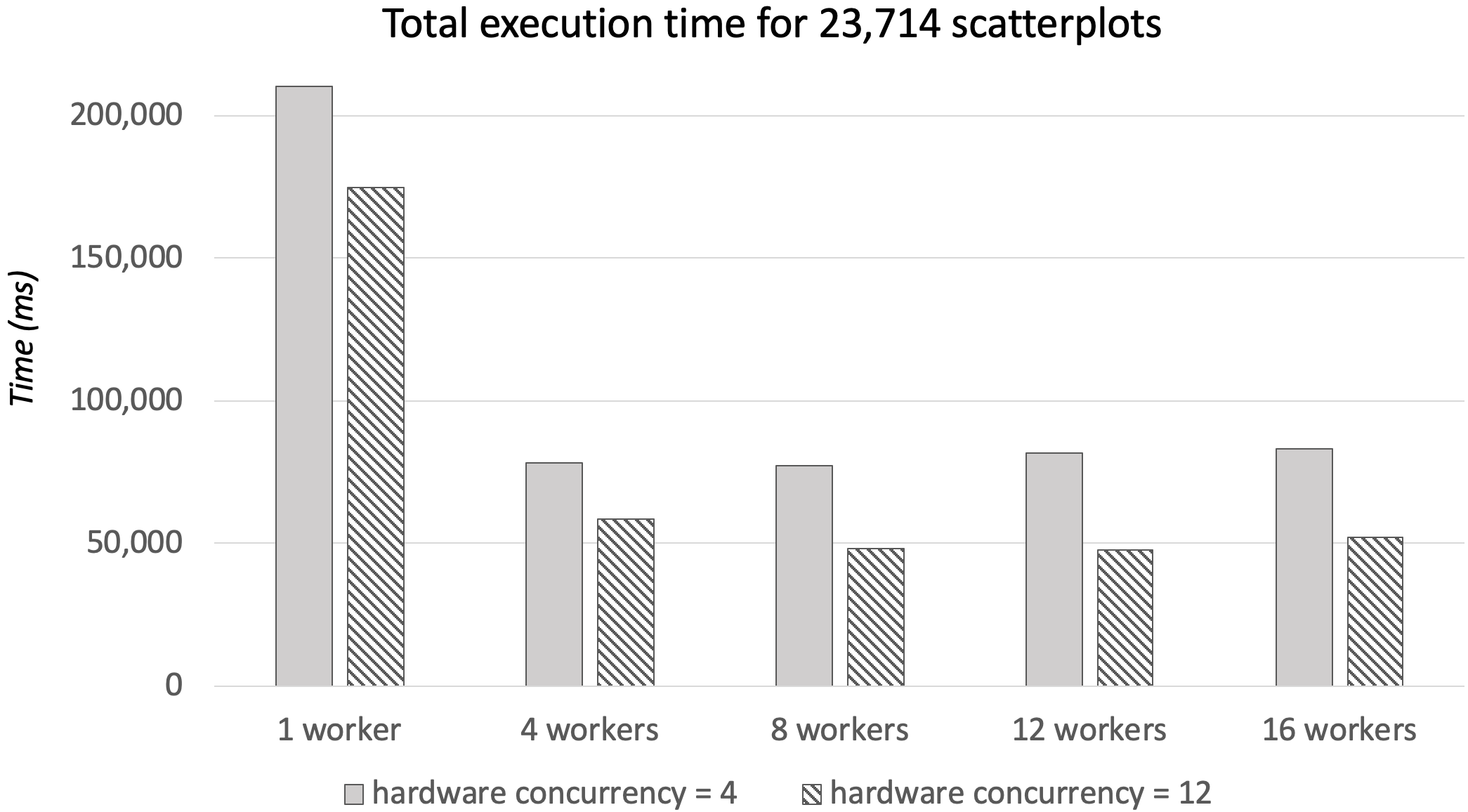}
 \caption{\label{fig:parallelexecutiontime} Computation times (in milliseconds) for 23,714 (original and \loo{} scatterplots) from datasets with various sizes in different parallelism configurations.}
\end{figure}

In this section, we evaluate the running times of \theName{} prototype with different settings to find the best parallelism configuration(s). The total of 23,714 (original and \loo{} scatterplots) from the datasets described in section \ref{sec:Datasets} were tested. These datasets were executed 30 times in each setting (to make sure the reported execution times are stable and not happened solely by chances due to the stochastic nature of computer execution time) then the averaged execution times are reported as in Figure \ref{fig:parallelexecutiontime}. All tests were performed on two computers, the first computer with 2.9 GHz Intel Core i5, macOS Mojave Version 10.14.3, 8 GB RAM, and hardware concurrency = 4 and the second one with 2.6 GHz Intel Core i7, macOS Mojave Version 10.14.3, 16 GB RAM, and hardware concurrency = 12. The parallelism configurations are 1, 4, 8, 12, and 16 web workers correspondingly. We explored these number of workers because they are around the number of hardware concurrency of the testing platforms (4, and 12 respectively).

Generally, using multiple workers helps to improve the computation time about three times and even more in devices with higher concurrency support. Also, the benefits of running more workers come with the cost of their creation and communication overheads, as of our experiment, a number of workers close or equal to the hardware concurrency support will give the best performance. The hardware concurrency support for different devices and operating systems could be determined in JavaScript as $navigator.hardwareConcurrency$, and this is the default parallelization setting of our system.

\subsection{Extending to higher-dimensional data}
The concepts discussed in this paper are specific to two dimensional (2D) temporal datasets. However, they could be generalized to detect outliers for a higher number of dimensions (nD) time series with a few modifications. In term of outlying score computation, we need to extend the Euclidean distance calculation from 2D to nD. In the nD version, we could also explore other options for distance calculation to avoid the ``curse of dimensionality". For instance, the Manhattan distance metric ($L_1$ norm) might be preferable than the Euclidean distance metric ($L_2$ norm), or even the $L_k$ norm where $k$ is a fraction should be explored~\cite{Anand2009Linfinity}. The visualization components and interactive operations of our prototype remain valid, except that the radar-charts can be used to replace scatterplots. 

\section{Conclusion}
In this paper, we have proposed a new approach for visualizing the outlying temporal profile of each data entry with respect to the overall distributions in two-dimensional temporal datasets and also discussed the extension of the ideas to a higher number of dimensions. Our approach is based on the leave-one-out strategy for measuring the significance of individual data points in computing outlying as a whole. This approach not only allows us to detect multivariate outliers but avoid both masking and swamping effects. We demonstrated our \theName{} prototype on various use cases of the US employment data, social and economic data from the World Bank database, and health status of high-performance computing systems. We also evaluated computing times to provide users with an idea of how long it takes to use our approach for certain datasets. The running time evaluations prove that our approach can scale well with large data thanks to binning, redundant checking before performing outlying computation on the leave-one-out plots, and the use of multiple web workers.

\bibliographystyle{abbrv-doi}

\bibliography{template}
\end{document}